\documentclass[prl,twocolumn,preprintnumbers,super,citeautoscript]{revtex4}
\usepackage[english]{babel}
\usepackage[T1]{fontenc}
\usepackage[latin1]{inputenc}
\usepackage[]{amsmath,amsfonts,amssymb,bm}
\usepackage{array,dcolumn,delarray,hhline,tabularx,longtable}
\usepackage{afterpage,xspace,latexsym,layout}
\usepackage[]{graphicx}
\usepackage[dvipdfm,usenames]{color}
\usepackage[dvipdfm,colorlinks,
	linkcolor=Violet,linkcolor=Violet,anchorcolor=Violet,citecolor=Violet,
		filecolor=Violet,menucolor=Violet,pagecolor=Violet,urlcolor=Violet
	]{hyperref}
\newcommand{\dd}{\text{d}}		

\begin{document}
\title{Invariants of free turbulent decay}
\author{Antoine Llor}
\email{antoine.llor@cea.fr}
\affiliation{\mbox{CEA, Commissariat à l'Energie Atomique, Direction des Programmes,}
\\ Bât.~Siège, 91191 Gif-sur-Yvette Cedex, France}
\date{November 13, 2006}
\begin{abstract}

	In practically all turbulent flows, turbulent energy decay is present and competes with numerous other phenomena. In Kolmogorov's theory, decay proceeds by transfer from large energy-containing scales towards small viscous scales through the ``inertial cascade.'' Yet, this description cannot predict an actual \emph{decay rate}, even in the simplest case of homogeneous isotropic turbulence (HIT). As empirically observed over 50 years, the steepness of the ``infrared'' spectrum---at scales larger than energy-containing eddies---determines decay, but theoretical understanding is still missing. Here, HIT decay laws are derived from angular momentum invariance at large scales---an approach first mentioned by Landau in 1944, but unduly dismissed later. This invariance also predicts the so-far unexplored turbulence decay in layer, tube, and spot zones in a fluid at rest. Beyond expanded and simplified theoretical descriptions, these findings suggest new practical modeling strategies for turbulent dissipation, often deficient in applied simulations.
\end{abstract}
%
\maketitle
%
\paragraph*{Background.}

	Incompressible homogeneous isotropic turbulence (HIT) at high Reynolds number evolves under the sole influence of its own dissipation: no external source of a characteristic quantity exists (flow size or velocity, energy production, etc.). After some initial transients, it is thus expected to reach a self-similar behavior whereby, with the proper origin on time $t$,
\begin{align}\label{eq:kl}
k &\propto t^{-n},
&
\lambda &\propto t^{1-n/2},
\end{align}
$k$ being the (per mass) mean turbulent kinetic energy, and $\lambda$ the integral length scale---the typical size of the large energy-containing turbulent eddies. As first noted by Kolmogorov \cite{Kolmogorov}, self-similarity of decay reflects the existence of an invariant in the turbulent field, which can be written as
\begin{align}
I &= k\lambda^m,
&
\text{and thus~} n &= \frac{2m}{2+m},
\end{align}
according to Eqs.~(\ref{eq:kl}). Experiments (wind tunnels, water channels, cryogenic helium, etc.) provide values with a clear clustering around $n\approx1.2$ \cite{Skrbek,Antonia} corresponding to $m\approx3$.

\begin{figure}
\includegraphics[width=2.5in]{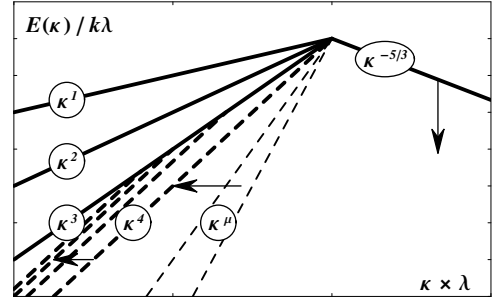}
\caption{\label{fig:SD}Schematic spectral density profiles of energy for HIT (in logarithmic coordinates scaled by the time-dependent energy $k$ and the integral length scale $\lambda$). Depending on its logarithmic slope $\mu$, the infrared range (i)~transitions quickly towards $\mu=4$ for $\mu<4$, (ii)~slowly evolves towards $\mu\approx3$ for $3\lesssim \mu\leq4$, or (iii)~is invariant for $\mu\lesssim3$, yielding self-similar decay of HIT. The inertial range decays but preserves its Kolmogorov $-5/3$ slope.}
\end{figure}

	This simple dimensional analysis of mean values is not sufficient to characterize neither the underlying random velocity field $\bm{u}(\bm{r},t)$, nor the fluid dynamics process which produces an invariant starting from a given initial condition $\bm{u}(\bm{r},0)$. The next descriptive order is provided by the velocity correlation tensor $\overline{u_i(\bm{0},t)u_j(\bm{r},t)}$ (the overbar stands for ensemble averaging). With the more common representation in Fourier space, it can be reduced to the spectral density of turbulent kinetic energy $E(\kappa)=\int\,[\int\frac{1}{2}\,\overline{u_i(\bm{0})u_i(\bm{r})}\,e^{i\bm{\kappa\cdot r}}\dd^3\bm{r}]\,\bm{\kappa}^2\dd^2\omega_{\bm{\kappa}}$ thanks to the conditions of average homogeneity and isotropy, and thus $k=\int E(\kappa)\dd\kappa$. In fully developed turbulence $E(\kappa)$ peaks around $\kappa_\lambda\approx\pi/\lambda$ (see Fig.~\ref{fig:SD})
and follows Kolmogorov's scaling $E(\kappa)\sim\kappa^{-5/3}$ in the inertial range---from $\kappa_\lambda$ up to Kolmogorov's dissipation wave number.

	The infrared range of $E(\kappa)$---below $\kappa_\lambda$---was very early found to be related to the invariant $I$ \cite{Loitsyanskii,Batchelor,Saffman,Davidson}, but it is difficult to produce and control in experiments. Only recently could its influence be thoroughly observed by numerical means\cite{Lesieur,Ossia}, showing three different regimes depending on the steepness of its self-similar $\kappa^\mu$ profile (see Fig.~\ref{fig:SD}):

\noindent (i)~For $\mu>4$, a quick transition towards $\mu=4$ takes place, typically in a few turnover times $\lambda/\sqrt{k}$. This well-documented process\cite{Batchelor} designated as ``backscattering'' is due to the coupling of wave numbers above $\kappa_\lambda$ into elongated triads through the non-linear term of the Navier--Stokes equation.

\noindent (ii)~For $3\lesssim \mu\leq4$---including $\mu=4$ as resulting from point (i) above,---a slow, non-self-similar asymptotic evolution towards $\mu\approx3$ takes place. This produces two close but different regimes within the infrared range (see Fig.~\ref{fig:SD}), but an almost self-similar evolution of $k(t)$ can eventually be observed for late-enough times.

\noindent (iii)~For $\mu\lesssim3$---including $\mu\approx3$ as resulting from point (ii) above at very long times,---a full self-similar decay of the spectrum is obtained, i.e. $E(\kappa)$ is invariant with respect to the reduced units $k$, $\lambda$ and their combinations. Moreover, the infrared profile is fully \emph{invariant} in ordinary units---this is the so-called ``permanence of large eddies''---and the overall invariant of turbulent decay is thus given by $m=\mu+1$. The initial preparation of the infrared spectrum of HIT thus defines the invariant. As explained by Saffman \cite{Saffman}, $m=3$ or $\mu=2$ prevails in most experiments if they all tend to produce randomly distributed initial impulses on the energy containing structures in the field.

	In summary, it is well understood how an \emph{assumed} permanence of large eddies produces an invariant \cite{Batchelor,Saffman,Davidson}---ultimately controlling $n$,---how an initial $\kappa^2$ infrared profile is generated \cite{Saffman}, and how an initial $\kappa^\mu$ profile with $\mu>4$ is quickly reduced to $\mu=4$ \cite{Batchelor}. However, a formal \emph{proof} of the permanence of large eddies for $\mu\lesssim3$ does not seem to have been produced so far \cite{Batchelor,Lesieur,Ossia}. The present article is focused on this critical aspect and its extension to situations other than HIT.

	Incidentally, one should notice that in the present work the far infrared and dissipative ranges---at $\kappa$ typically below or above the ranges represented in Fig.~\ref{fig:SD}---are not constrained to evolve in a self-similar manner, because they marginally affect the overall behavior of energy containing scales. In contrast, the infrared contribution to $k$ is by no means negligible: not only does it control the decay rate, but it also represents about $\frac{2}{3}/(\mu+1)$ of the energy contained in the inertial range, a substantial 22\% for $\mu=2$. 
%
\paragraph*{Landau's invariants.}

	The first and most direct relationship between the infrared spectrum and a decay invariant was postulated by Loitsyanskii \cite{Loitsyanskii} in direct space, using the normalized longitudinal space-correlation function of velocity $f(s,t)=\overline{u_z(\bm{0},t)\,u_z(s\bm{\hat z},t)}\,/\,\overline{u_z^2(\bm{0},t)}$. $f(s)$ is related to $E(\kappa)$ by Fourier transform, and thus the inertial and infrared (or large scales) ranges correspond respectively to $f(s)\approx1-(s/\lambda)^{2/3}$ for $s<\lambda$, and $f(s)\propto(s/\lambda)^{-m}$ for $s>\lambda$. Thus $f(s)\sim s^{-3}$ at large scales for Saffman's $E(\kappa)\sim\kappa^2$ infrared spectrum.

	The evolution of $f(s,t)$ is given by the Kármán--Howarth equation which is deduced from the Navier--Stokes equation and involves a normalized two-point triple-correlation function $K(s)$. In space integrated form it yields the invariance of Loitsyanskii's integral \cite{Loitsyanskii}
\begin{subequations}\label{eq:ILoi}\begin{align}
I_\text{Loi.} &= k \int_0^\infty f(s,t) \, s^4 \dd s,
\\
		&\propto k(t)\lambda^5(t)
			\text{~~~~if~} f(s,t)=f\bm{(}s/\lambda(t)\bm{)},
\end{align}\end{subequations}
provided that $f(s)$ and $K(s)$ decrease faster than $s^{-5}$. When it was found \cite{Batchelor,Saffman} that $f(s)$ and $K(s)$ actually decrease \emph{slower} than $s^{-5}$ in all practical situations---as mentioned in the introductory paragraphs---Loitsyanskii's approach was simply dismissed.

	Yet, when $I_\text{Loi.}$ was still considered invariant, Landau suggested a particularly appealing physical interpretation \cite{Landau,Davidson}. From dimensional and scaling arguments, he remarked that Loitsyanskii's integral was similar to the per-volume angular-momentum variance of an arbitrarily large spherical volume $\varOmega(R)$, which he estimated to grow as $R^3$: $I_\text{Loi.}\propto\lim\overline{H^\varOmega_zH^\varOmega_z}/R^3$ (no sum on $z$) as $R\rightarrow\infty$, where $\bm{H}=\int_\varOmega \bm{r \times u}\,\dd^3\bm{r}$. As he also estimated the average torque $T^\varOmega_z(R)$ on $\varOmega(R)$ to grow as $R^2$ only, the invariance of $I_\text{Loi.}$ was recovered. 
However, Landau did not develop his argument in full, and, although it appears convincing, it could as well lead to $\overline{H^\varOmega_zH^\varOmega_z}\sim R^5$ instead of $R^3$ by a similar scaling analysis: each point yields a typical $kR^2$ contribution to $\overline{H^\varOmega_zH^\varOmega_z}$ and the superposition of uncorrelated contributions in the double integral yields a further $R^3$.

	Landau's interpretation actually reduces to solving the stochastic differential equation $\dd_tH^\varOmega_z(R)=T^\varOmega_z(R)$---similar to Langevin's equation for Brownian motion---and seek invariance conditions. The explicit calculation (see Supplementary Methods, §§1--3) eventually yields neither of the expected results as
\begin{subequations}\label{eq:HT}\begin{align}
\label{eq:HHTT}
& \dd_t \overline{H^\varOmega_zH^\varOmega_z}
		\sim (\lambda/\sqrt{k}) \, \overline{T^\varOmega_zT^\varOmega_z}
			\text{~~~~for~} R\rightarrow\infty,
\\
\label{eq:HHI}
& \overline{H^\varOmega_zH^\varOmega_z}
		= R^4 k \int_0^{2R} P(s/R) \, f(s,t) \, s^3 \dd s,
\\
\label{eq:TTI}
& \overline{T^\varOmega_zT^\varOmega_z}
		= R^4 k^2 \int_0^{2R} \sum_\alpha
				P_\alpha(s/R) \, f_\alpha(s,t) \, s\, \dd s,
\end{align}\end{subequations}
where $f_\alpha$ are the five independent normalized two-points, quadruple space-correlation functions of velocity; and $P$ and $P_\alpha$ are even polynomials such that $P_{(\alpha)}(0)\neq0$ and $P_{(\alpha)}(2)=0$. In the derivation, a scaling of $\overline{H^\varOmega_zH^\varOmega_z}$ in $R^5$ appears as expected but it is eventually reduced to $R^4$ by the incompressibility condition on $f$. Also noticeable are the vanishing contributions from pressure fluctuations in the torque due to the symmetry of $\varOmega$.

	In general, $\overline{H^\varOmega_zH^\varOmega_z}$ should not be invariant at large scales when $R\rightarrow\infty$, as $\overline{T^\varOmega_zT^\varOmega_z}$ also appears to scale as $R^4$. However, as with Loitsyanskii's integral, the behaviors of $f(s)$ and $f_\alpha(s)$ when $s\rightarrow\infty$ must be taken into account in analyzing the scalings with $R$ in Eq.~(\ref{eq:HT}), since they can produce divergent integrals: for instance, with Saffman's $f(s)\sim s^{-3}$, $\overline{H^\varOmega_zH^\varOmega_z}\sim R^5$ as first expected intuitively. For a generic $f(s)\sim s^{-m}$ behavior at $s\rightarrow\infty$, it is thus found that
\begin{subequations}\label{eq:I}\begin{align}
\label{eq:II}
I_\text{Lan.} &= \lim_{R\rightarrow\infty} \Big[\, R^{m-4} \, k
		\int_0^{2R} P(s/R) \, f(s,t) \, s^3 \dd s \,\Big]
\\
\label{eq:IL}
		&\propto k(t)\lambda^{m}(t)
			\text{~~~~if~} f(s,t)=f\bm{(}s/\lambda(t)\bm{)}
\end{align}\end{subequations}
is a finite quantity for $m<4$. Now, because in general $f_\alpha(s) \sim f^2(s)\sim s^{-2m}$ at $s\rightarrow\infty$, $\overline{T^\varOmega_zT^\varOmega_z}$ always scales as $R^4$ for $1<m<4$, and $\overline{H^\varOmega_zH^\varOmega_z}$ does become asymptotically invariant at $R\rightarrow\infty$ for such values of $m$---or for $0<\mu<3$ in spectral space. This proves the so-far empirical conditions for the permanence of large eddies reviewed in the introductory paragraphs.

	At this point the significance of the invariant as given in Eqs.~(\ref{eq:I}) requires some elaboration:
	(i)~Although the \emph{physical} invariant is always the same---the angular momentum variance,---the \emph{expression} of the invariant also depends on an \emph{external} parameter $m$ which is not specified a priori: the long-range behavior of $f(s)$---or the infrared behavior of $E(\kappa)$.
	(ii)~A given initial long-range behavior of $f(s)$ will be preserved at later times according to Eq.~(\ref{eq:II}), i.e. $m$ is invariant, if $m<4$---or $\mu$ if $\mu<3$.
	(iii)~Furthermore, if the evolution (by the Navier-Stokes equation) preserves the self-similarity of $f(s,t)$, then $k\lambda^m$ is also invariant according to Eq.~(\ref{eq:IL}).
	(iv)~As a corollary, because evolution does not affect the infrared spectrum below the integral length scale, and ensures a quasi-equilibrium self-similar profile in the inertial range, it is necessary and sufficient to have \emph{initially} a self-similar profile at large scales, such as $f(s)\sim (s/\lambda)^{-m}$ with $m<4$ to ensure a $k\lambda^m$ invariant.
	(v)~For initial conditions at $m>4$ the inertial range dominates the integral in Eq.~(\ref{eq:II}), and $\overline{H^\varOmega_zH^\varOmega_z}$ is not asymptotically invariant according to Eq.~(\ref{eq:HHTT}). A universal behavior, i.e. independent of the initial infrared spectrum profile but not necessarily self-similar, must then appear and eventually converge towards $m\approx4$ and $n\approx4/3$. This is the maximum decay rate of turbulence.

	It must be stressed here that a rescaling of Loitsyanskii's integral with the proper $R$ power as for $I_\text{Lan.}$ cannot produce the same invariants of HIT evolution because different integration weights are involved---$s^4\dd s$ in Eq.~(\ref{eq:ILoi}) instead of $s^3\dd s$ in Eq.~(\ref{eq:II}). In particular, the maximum decay rate at $n\approx4/3$ cannot be predicted. This reflects a profound difference between the approaches of Loitsyanskii and Landau---despite the latter's incidental claim of equivalence \cite{Landau}---which can be traced to alternative integrations of the stochastic equation $\dd_tH^\varOmega_z(R)=T^\varOmega_z(R)$ (see Supplementary Methods, §1): 
Loitsyanskii's approach leads to $\dd_t\overline{H^\varOmega_zH^\varOmega_z}=\overline{T^\varOmega_zH^\varOmega_z}$ instead of Eq.~(\ref{eq:HHTT}), and thus produces the two-point \emph{triple}-correlation of the Kármán--Howarth equation instead of the two-point \emph{quadruple}-correlations of Eq.~(\ref{eq:TTI}).
%
\paragraph*{Impulsive approach to Landau's invariants.}

\begin{figure}
\includegraphics[width=1.2in]{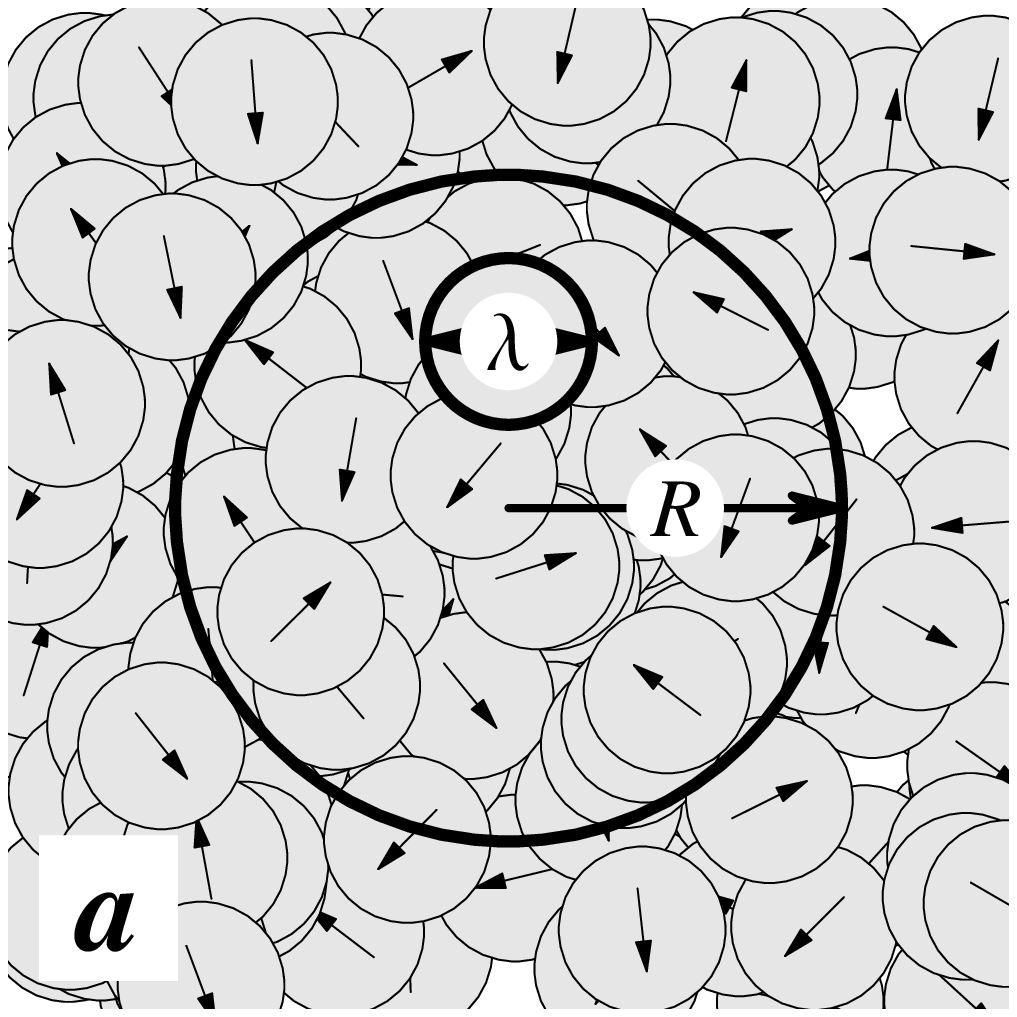}
\rule{.3in}{0in}
\includegraphics[width=1.2in]{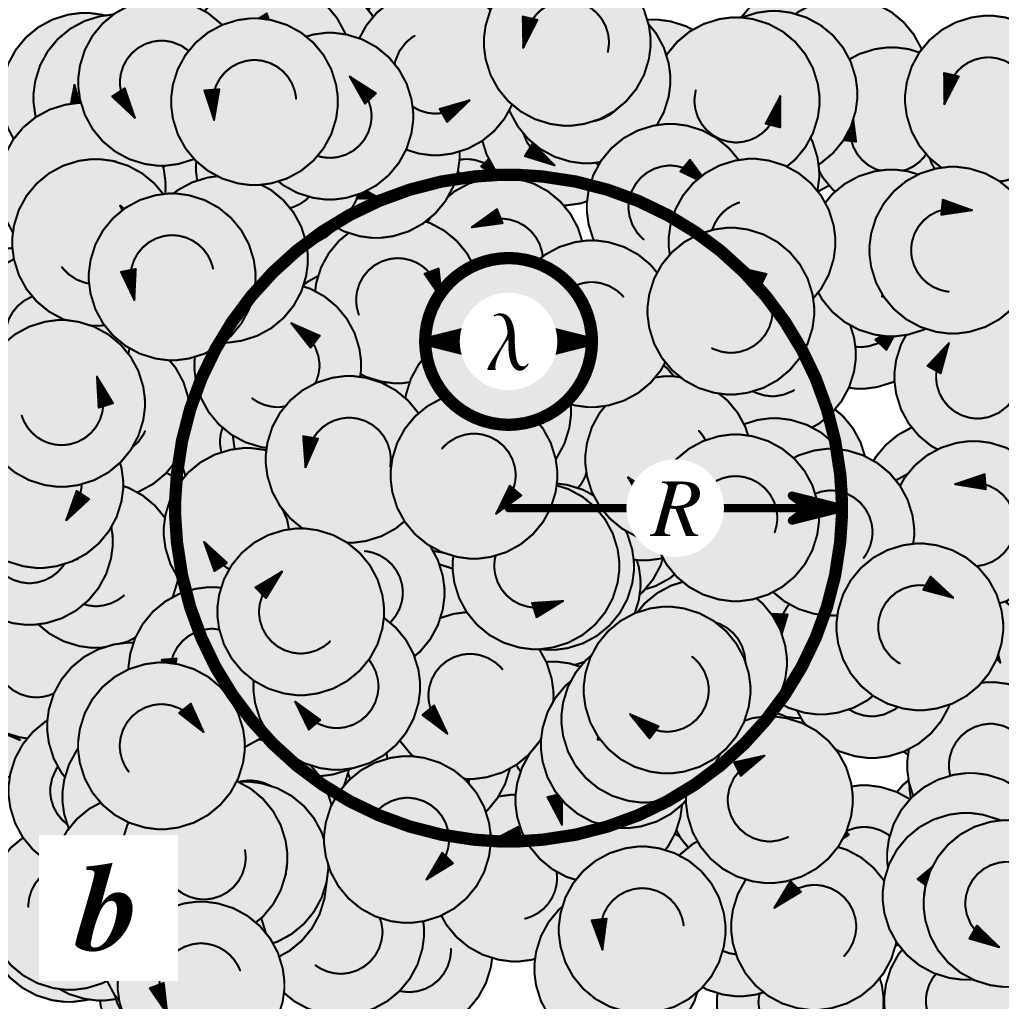}
\caption{\label{fig:SB}Schematic representation of large turbulent eddies in HIT (shaded circles) moving randomly with typical velocities $\sim\sqrt{k}$ in two basic modes: \textbf{a}) translations or Saffman correlation, $\overline{H^2}\sim R^5k\lambda^3$, \textbf{b}) rotations or Batchelor correlation, $\overline{H^2}\sim R^3k\lambda^5$.}
\end{figure}

	In the preliminary scaling analysis leading to $\overline{H^\varOmega_zH^\varOmega_z}\sim R^5$, the velocity field was implicitly assumed to display the structure represented in Fig.~\ref{fig:SB}a.
Landau's result $I_\text{Lan.}\sim R^3$ would actually correspond to Fig.~\ref{fig:SB}b, where large eddies have no \emph{translational} motion, and just \emph{rotational} motion, each contributing by $k\lambda^2$ instead of $kR^2$ to the angular momentum variance.

	The representations in Fig.~\ref{fig:SB}, however simplistic they might appear, are actually well suited to carry out a scaling analysis and to derive the $R$ dependence of the invariants. Indeed the scaling behavior is marginally sensitive to the details of (i)~the motions \emph{within} the large eddies, contributing to lower powers of $R$, and (ii)~the interactions and correlations \emph{between} neighboring structures. The last point stems from Saffman's projection procedure \cite{Saffman} which generates a \emph{kinematically consistent} velocity field $\bm{u}$ from any given \emph{arbitrary impulse} field $\bm{i}$
\begin{align}\label{eq:SafProj}
\bm{u} &= \bm{i} - \bm{\nabla} q
&
\text{with~} \Delta q &= \bm{\nabla\!\cdot i},
&
\text{thus~} \bm{\nabla\!\cdot u} &= 0.
\end{align}
Accordingly, \emph{Fig.~\ref{fig:SB} represents impulse fields} from which kinematically consistent velocities can be produced. Now, a crucial property of Saffman's procedure is that \emph{it strictly preserves the angular momentum of any spherical volume} $H^\varOmega_z$ since, here again, the symmetry cancels the torque due to the ``impulse pressure forces'' $\bm{\nabla} q$.

	The torque $T^\varOmega_z$ is not preserved by Saffman's procedure, but the $R^4$ behavior of $\overline{T^\varOmega_zT^\varOmega_z}$ is still recovered from a scaling analysis on the impulsive representation, as given in Supplementary Methods, §4. In particular, the differences of short range velocity correlations between Figs.~\ref{fig:SB}a and \ref{fig:SB}b do not impact the $R^4$ factor: in $T^\varOmega_z$'s expression, as the velocity field of the turbulent structures is just \emph{sampled at the surface} $\partial\varOmega$, possible details of the velocity correlation functions are lost. 
The basic $R^4$ scaling derived from Fig.~\ref{fig:SB}a is then valid for all cases: according to Eq.~(\ref{eq:TTI}), only with strongly long-range-correlated velocities can the torque's scaling be changed---i.e. if $f_\alpha(s)$ decays slower than $s^{-2}$ or $f(s)$ slower than $s^{-1}$. Rigorous and physically consistent conclusions on both the scaling and the time dependence of Landau's invariants can thus be drawn from the present simple impulsive representation.

	As previously shown \cite{Saffman}, Saffman's projection generally preserves the $\kappa^\mu$ profile of the infrared spectrum: in the canonical case of Fig.~\ref{fig:SB}a first considered by Saffman \cite{Saffman}, random translations produce $E(\kappa)\sim\kappa^2$. At the same time however, it can dramatically modify the long range profiles of the two-point self-correlations $f$ of $\bm{i}$ and $\bm{u}$ \cite{Saffman}: again in Fig.~\ref{fig:SB}a, an arbitrary $\bm{i}$ representing random long-range-uncorrelated translations---thus with transcendentally vanishing $f_i(s)$---eventually yields a long-range-correlated $\bm{u}$ with an algebraically vanishing $f(s)\sim s^{-3}$---reflecting the laminar and dipolar character of long range correlations steming from Eq.~(\ref{eq:SafProj}). This last result substituted in Eq.~(\ref{eq:HHI}) recovers the proper behavior $I_\text{Lan.}\sim R^5$ as estimated from the impulse field $\bm{i}$.

\begin{figure}
\parbox[c]{.5in}{\includegraphics[width=.5in]{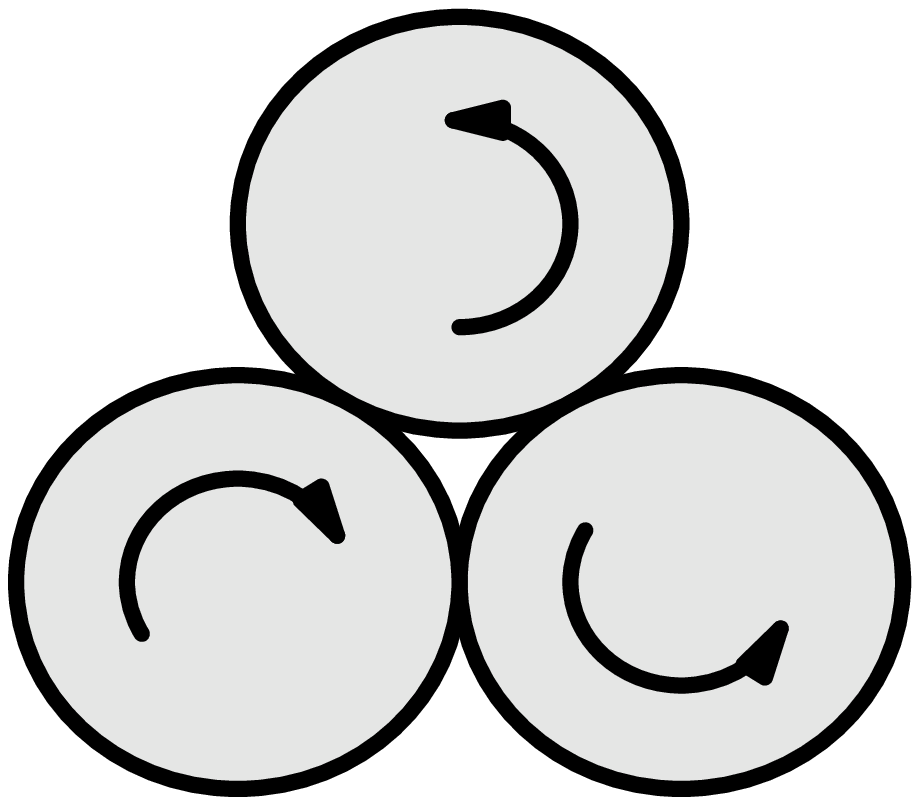}}
$\longrightarrow$
\parbox[c]{.5in}{\includegraphics[width=.5in]{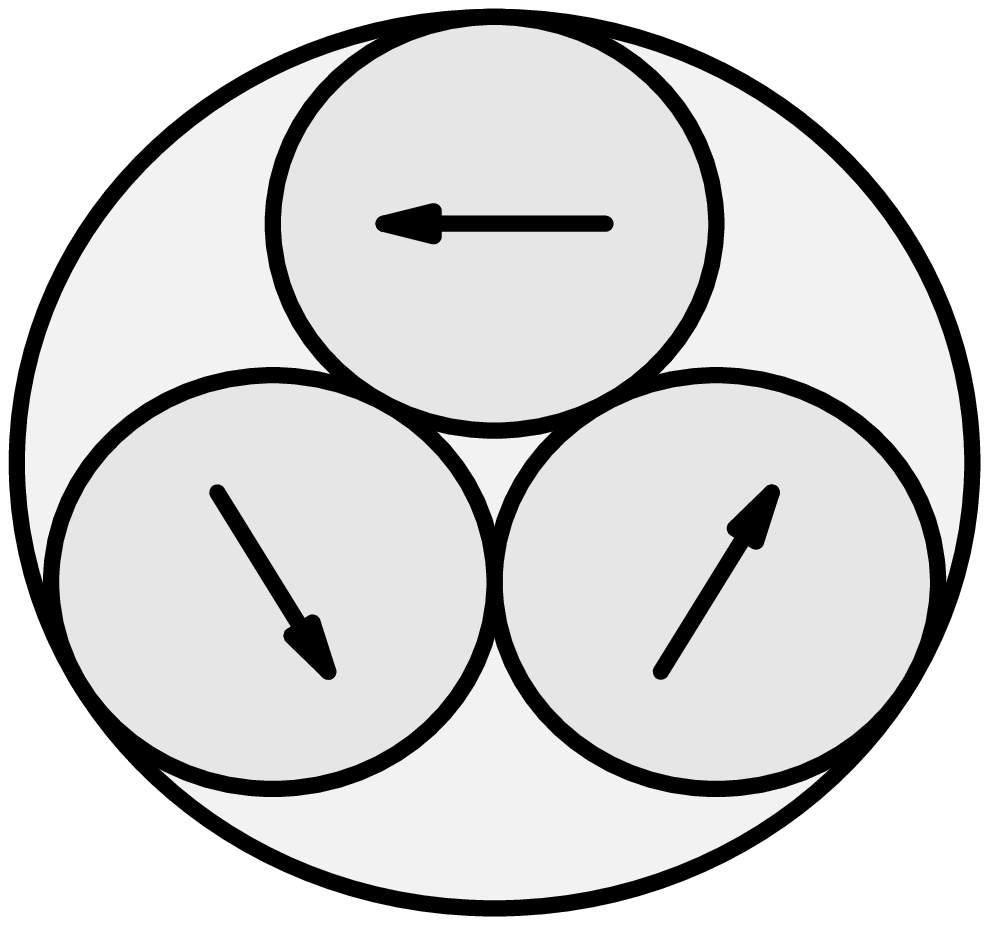}}
\caption{\label{fig:BckSct}Impulse representation of large eddies undergoing so-called ``backscattering'' in Batchelor type HIT. The process, albeit slower, also affects ``big frozen eddies'', much larger than the integral length scale.}
\end{figure}

	The $E(\kappa)\sim\kappa^4$ infrared profile first considered by Batchelor \& Proudman \cite{Batchelor} corresponds to a purely rotational impulse field as in Fig.~\ref{fig:SB}b: it is indeed found $I_\text{Lan.}\sim R^3$ in this case, both with the impulsive scaling analysis and with Eq.~(\ref{eq:II}). Since $m>4$, the expected non-self-similar evolution is obtained where, akin to the collisionnal processes at molecular scales, the turbulent energy is in part redistributed on translational degrees of freedom with lower $m\lesssim4$---a well-known ``backscattering'' phenomenon illustrated with an impulse image in Fig.~\ref{fig:BckSct}.
However, the typical turnover rate at scale $R$ is given by $(\overline{H^\varOmega_zH^\varOmega_z})^{1/2}/(\rho R^{5})\sim R^{-2}$ and at very large scales the redistribution is practically frozen.

	Saffman's impulse field appears as most likely when considering turbulence generated by experimental set ups (as grids) devoid of specific devices which could constrain correlations such as Batchelor's \cite{Saffman}. However, (i)~in many numerical simulations, initial conditions are implicitly of Batchelor type, and (ii)~any experimental or numerical setup has an upper bound for $R$ and produces Batchelor fields at these largest scales. Other types of initial conditions seem unrealistic or highly specific: for instance $E(\kappa)\sim\kappa$ for $\kappa\rightarrow0$ ensures full self-similarity of the whole spectrum down to dissipative scales \cite{Speziale}, but no general physical mechanism is known to produce such a constraint.
%
\paragraph*{Invariants in layer and tube.}

\begin{figure}
\includegraphics[width=3.375in]{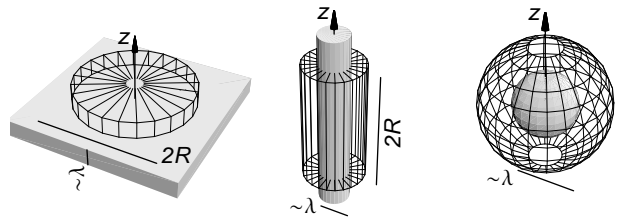}
\caption{\label{fig:2D1D0D}Volumes $\varOmega(R)$ (framing lines) displaying asymptotically invariant variance of angular momentum for turbulent layer, tube, and spot (shaded volumes).}
\end{figure}

	The impulsive approach to turbulence invariants has the important property of being easily extended to the less trivial situations of self-similar layers and tubes of turbulence. Only the $z$ component of angular momentum along the $C_\infty$ symmetry axis is now preserved: the previous derivation for HIT is simply adapted to these cases taking the integration volumes sketched in Fig.~\ref{fig:2D1D0D}.
Details are provided in the Supplementary Methods, §4, but the basic steps and assumptions are quickly summarized below.

	Because of self-similarity, typical layer thickness or tube diameter scale as $\lambda$. As in HIT, for each motion mode (Saffman, Batchelor, etc.) of each geometry, factors in $\overline{H^\varOmega_zH^\varOmega_z}$ come from velocity fluctuations, typical distance to axis, and integration domain over correlated and uncorrelated distances---below and beyond $\lambda$. With vanishingly small contributions of velocity fluctuations in the surrounding inviscid laminar fluid, the torque is dominated by the parts of $\varOmega$'s surface which intersect the turbulent zone. 
Then, comparing behaviors as $R\rightarrow\infty$, $\overline{H^\varOmega_zH^\varOmega_z}$ is also found to be asymptotically invariant for $m$ below some maximum value. Finally, Saffman's projection procedure applies as well: impulse is confined to the turbulent zone, velocity extends into laminar regions, but angular momentum is preserved by the projection. Resulting invariants and evolution exponents are collected in Table~\ref{tab:inv} for Saffman impulse fields, and for correlations corresponding to maximum decay rates.
\begin{table*}
\caption{\label{tab:inv}Turbulent relaxation in geometries of Fig.~\ref{fig:2D1D0D}: invariants and evolution exponents (S. \& B. = Saffman \& Batchelor).}
\begin{ruledtabular}
\begin{tabular}{l*{7}{c}}
Geometry
	& \multicolumn{2}{c}{HIT}
	& \multicolumn{2}{c}{Layer (IRM)}
	& \multicolumn{2}{c}{Tube}
	& \multicolumn{1}{c}{Spot}
		\\ \cline{2-3} \cline{4-5} \cline{6-7} \cline{8-8}
\rule[0em]{0em}{1.1em}Invariant type
	& Saffman & Maximum
	& Saffman & Maximum
	& S. \& B. & Spin
	& Spin \\ \hline
\rule[0em]{0em}{1.1em}$m$ (invariant $k\lambda^m$)
	& 3 & 4
	& 4 & 5
	& 7 & 6
	& 8 \\
$n$ (decay exponent of $k$)
	& 6/5 & 4/3
	& 4/3 & 10/7
	& 14/9 & 3/2
	& 8/5 \\
$1-n/2$ (growth exponent of $\lambda$)
	& 2/5 & 1/3
	& 1/3 & 2/7
	& 2/9 & 1/4
	& 1/5 \\
$C_{\varepsilon2}$ ($\varepsilon$ model constant)
	& 11/6 & 7/4
	& 2 & 15/8
	& 19/10 & 2
	& 2
\end{tabular}
\end{ruledtabular}
\end{table*}
The strength of the impulsive approach is now apparent as all these new non-trivial results are readily obtained ``on the back of the envelope,'' without any detailed knowledge of the complex correlation functions that describe the velocity field in these anisotropic and inhomogeneous flows.

	As reported in Table~\ref{tab:inv}, Batchelor and Saffman invariants happen to coincide for the tube of turbulence because both are produced by identical $k\lambda^2$ as point contributions to $\overline{H^\varOmega_zH^\varOmega_z}$. The case of a spot of turbulence is also considered but, in contrast to the three previous situations, the angular momentum cannot fluctuate: mean angular momentum due to the residual overall spinning then provides the invariant in Table~\ref{tab:inv}. The turbulent tube is also compatible with mean overall spinning---centrifugal forces remain finite---with the corresponding invariant in Table~\ref{tab:inv}. The relaxation of turbulence for spinning zones is constrained by these invariants for as long as mean angular momentum dominates other terms---such as angular momentum variance in the tube case.

	Besides HIT, experimental and numerical results appear to be available for the turbulent layer only, a situation also known as the incompressible Richtmyer--Meshkov instability (IRM) at vanishing Atwood number \cite{Dimonte,Shvarts}. The value of the IRM self-similar growth rate is still a matter of debate due to many perturbation sources affecting accuracy (limited time span and space extension, poorly defined or controlled initial conditions, etc.). However, up to reported uncertainties, values measured for $1-n/2$ do fall within the expected [2/7,1/3] range, with results from experiments and from numerical simulations closer respectively to the upper 1/3 and lower 2/7 bounds. 
This is consistent with the corresponding initial conditions: expected Saffman type in experiments and explicit Batchelor type in simulations---in the latter, calculations start from superpositions of random interfacial perturbations in the \emph{laminar linear regime}, hence with vanishing $H^\varOmega_z$. A previous theoretical prediction based on an implicit assumption of Saffman type turbulence also yielded $1-n/2=1/3$ as expected, but with a completely different approach which could be extented in 2D \cite{Inogamov}.

	The present predictions for different flow topologies---of higher practical relevance than HIT---provide guidance in the modeling of the equation of turbulent dissipation $\varepsilon$ \cite{Pope}. For instance, $\varepsilon$ is usually modeled as in the $k$--$\varepsilon$ model which, for the four situations considered here, can be approximately but accurately reduced to the bulk ODEs \cite{Llor}
\begin{subequations}\label{eq:ke}\begin{align}
\dd_t K &= - d \, (\dd_t\varLambda/\varLambda) K - E
\\
\dd_t E &= - d \, (\dd_t\varLambda/\varLambda) E - C_{\varepsilon2} E^2/K
\end{align}\end{subequations}
where $K$, $E$, and $\varLambda$ are the bulk averages of $k$, $\varepsilon$ and $\lambda$ over the turbulent zones, $\dd_t\varLambda/\varLambda$ terms describe the dilution effects in place of the flux terms of the PDEs, and $d=0$, 1, 2, or 3 for HIT, layer, tube or spot respectively. Self-similar solutions of Eq.~(\ref{eq:ke}) match the invariant based solutions if the model constant is
\begin{equation}\label{eq:CE2}
C_{\varepsilon2} = 3/2 + (1+d/2)/(m-d)
\end{equation}
yielding the values in Table~\ref{tab:inv}.

	Inspection of Table~\ref{tab:inv} shows that a single value of $C_{\varepsilon2}$---usually set to 1.92 \cite{Pope}---can never capture the three Saffman cases at once (let alone the others), possibly explaining some practical difficulties such as up to 15\% errors in growth rates when trying to simultaneously capture plane and cylindrical jets with a single model \cite{Pope}. Significant improvements could thus be contemplated when adding model equations aimed at ``sensing'' the dimensionality of the turbulent zone $d$, and at ``defining'' the behavior of angular momentum variance at large scales $m$. 
Anyhow, these correction will remain around 15\% since the influences of $m$ and $d$ in Eq.~\ref{eq:CE2} almost compensate each other when varying the flow topology. Incidentaly, this noticeable property of Eq.~\ref{eq:CE2} explains in part the often disregarded fact that most models are robust with respect to the effective Knudsen number of turbulence: models can and are applied to situations where $\lambda$ is similar to the characteristic flow size $L$ without major malfunctions, although they are calibrated through $C_{\varepsilon2}$ to capture HIT where $\lambda\ll L$.

	The generally accepted perception of the significance and status of $\varepsilon$ in turbulence modeling \cite{Pope} should be reassessed in view of the present results. The initial definition as the mean small-scale viscous dissipation $\overline{\nu u_{i,j} u_{j,i}}$ is now often abandoned in favor of the spectral flux, which assumes some form of spectral quasi-equilibrium whereby $E(\kappa)\propto\varepsilon^{2/3}\kappa^{-5/3}$ in the inertial range. As shown here however, turbulence decay is generally controlled by the infrared spectrum \emph{regardless of the detailed profile in the inertial range}: 
the very same results would hold even with non-Kolmogorov spectra---which do appear for instance in the presence of helicity, buoyancy or electromagnetic effects. In fact, $\varepsilon$ is solely defined as the time derivative of $k$, the energy in the energy containing range. Thanks to the assumption of global self-similarity of the energy containing range (infrared + inertial), an integral length scale $\lambda$ can be defined, and then $\varepsilon$ is constrained to match $k^{3/2}/\lambda$.

	It should not be considered as surprising that the self-similar decay of turbulence is totally independent of the detailed processes within the inertial range. Indeed, this situation is comparable to the perfectly inelastic collision of two bodies: the overall conservation of momentum fully defines the final state, regardless of whatsoever details on the mechanisms which have actually removed or dissipated the excess energy.
%
\paragraph*{Conclusions.}

	The present work has provided a rigorous answer to the long-standing issue of the permanence of large eddies and their relationship to turbulence decay, using angular momentum invariance at large scales. This approach has the advantage of being both extendable to less trivial cases of turbulent layer, tube and spot, and reducible to a calculationally simple impulsive picture. Beyond their intrinsic value and novelty in the theoretical study of turbulence, the resulting predictions on decay rates open up new approaches and understanding for improving the modeling of turbulent dissipation in applied simulations.
%

%
\paragraph*{Acknowledgments.}

	The author thanks O.~Poujade, L.~Jacquin, and B.~Aupoix for enlightening discussions.
\begin{widetext}
\newpage
\section{Supplementary Methods}
%
\subsection{1.~Landau's angular momentum invariance at large scales}

	One starts with the equation of fluctuating velocity in a zero mean velocity field (uniform constant density and vanishing viscosity are assumed)
\begin{equation}
\rho \partial_t u_i + \rho ( u_i u_j )_{,j} + p_{,i} = 0.
\end{equation}
Angular momentum fluctuations along $z$ integrated over a fixed volume $\varOmega$ then yield
\begin{equation}
\int_\varOmega \epsilon_{zij} r_i
	\left[ \rho \partial_t u_j + ( \rho u_j u_k )_{,k} + p_{,j} \right]
		\dd^3 \bm{r} = 0,
\end{equation}
or after elementary algebra
\begin{equation}\label{eq:dtH}
\dd_t H^\varOmega_z = T^\varOmega_z,
\end{equation}
where the fluctuating angular momentum and torque are
\begin{subequations}\begin{align}
\label{eq:H}
H^\varOmega_z(t)
	&= \int_\varOmega \epsilon_{zij} r_i \rho u_j \dd^3 \bm{r},
\\
\label{eq:T}
T^\varOmega_z(t)
	&= - \int_{\partial\varOmega} \epsilon_{zij} r_i
		( \rho u_j u_k + \delta_{jk} p ) \, \sigma_k \dd^2 \bm{r},
\end{align}\end{subequations}
$\bm{\hat\sigma}$ being the unit vector normal to the surface, with components $\sigma_i$. For integration volumes $\varOmega$ considered in this work, the orientation of the normal vector element $\bm{\hat\sigma}\dd^2\bm{r}$ cancels the $z$-component of the torque produced by pressure $p$. Pressure is thus disregarded in all the following.

	A time evolution equation for the variance of $H^\varOmega_z$ can then be obtained by a direct substitution of the torque expression, and accordingly
\begin{equation}
\dd_t \overline{ H^\varOmega_z(t) H^\varOmega_z(t) }
	= 2 \; \overline{ T^\varOmega_z(t) H^\varOmega_z(t) }
	= -2 \int\!\!\!\!\int_{\partial\varOmega\times\varOmega}
		\epsilon_{zij} \epsilon_{zi'j'} r_i r'_{i'}
			\, \rho^2 \overline{ u_j(\bm{r}) u_k(\bm{r}) u_{j'}(\bm{r'}) }
				\, \sigma_k \dd^2 \bm{r} \, \dd^3 \bm{r'}.
\end{equation}
In this way, the two-point, triple space-correlation functions of velocity can be introduced, in close similarity with the usual Kármán--Howarth equation in volume integrated form. Now, this derivation, although rigorous and exact, is \emph{not} in the spirit of Landau's approach, which hinges critically on the argument that the torque becomes asymptotically negligible at large scales, and the angular momentum asymptotically constant (one is a surface integral and the other a volume integral). In this case one should instead derive an evolution equation of the angular momentum variance by integrating the stochastic differential equation (\ref{eq:dtH}) in the weak collision (or friction) limit---just as for Brownian motion. For asymptotically large $\varOmega$ the correlation time of $H^\varOmega_z(t)$ is much longer than that of $T^\varOmega_z(t)$, and one can thus write
\begin{equation}
\dd_t \overline{ H^\varOmega_z(t) H^\varOmega_z(t) }
	= 2 \; \overline{ T^\varOmega_z(t) \int_0^t T^\varOmega_z(t') dt' }
	\approx 2 \int_0^\infty
				\overline{ T^\varOmega_z(t) T^\varOmega_{z}(t-t') } dt'
	= 2 \; \overline{ T^\varOmega_z(t) T^\varOmega_z(t) } \; \tau(t),
\end{equation}
where the torque correlation time is defined and approximated as
\begin{equation}
\tau(t) = \frac{1}{\overline{ T^\varOmega_z(t) T^\varOmega_z(t) }}
	\int_0^\infty \overline{ T^\varOmega_z(t) T^\varOmega_{z}(t-t') } dt'
				\approx \lambda/\sqrt{k}.
\end{equation}
Here, it was assumed that the spinning associated with the angular momentum is small enough so as to produce negligible shear at the boundary: this avoids far-from-average velocity fluctuations which could significantly change the torque estimate and even produce a correlation between torque and angular momentum.

	In general, expanding $\overline{T^\varOmega_zT^\varOmega_z}$ shows that the time derivative of $\overline{H^\varOmega_zH^\varOmega_z}$ now depends on two-point, \emph{quadruple} space-correlation functions of velocity. Landau's approach is thus physically distinct from Loitsyanskii's, although eventual invariants may---and should---coincide in regimes where their respective approximations are compatible. Explicit expressions of $\overline{H^\varOmega_zH^\varOmega_z}$ and $\overline{T^\varOmega_zT^\varOmega_z}$ as integrals of velocity space-correlation functions are given in the next two sections for the simple case of HIT.
%
\subsection{2.~Angular momentum variance of a sphere in HIT}

	The angular momentum of a given spherical volume $\varOmega$ of radius $R$ around origin in a HIT field is given by Eq.~(\ref{eq:H}). Although it cancels on average, its average square norm does not and is given by
\begin{align}
\overline{H^\varOmega_iH^\varOmega_i}
	&= \int\!\!\!\!\int_{\varOmega^2}
		\epsilon_{ijk} \epsilon_{ij'k'} r_j r'_{j'}
			\, \rho^2 \overline{ u_k(\bm{r}) u_{k'}(\bm{r'}) }
				\, \dd^3 \bm{r} \, \dd^3 \bm{r'}
\notag \\
	&= \rho^2 \int\!\!\!\!\int_{\varOmega^2}
		(\delta_{jj'}\delta_{kk'}-\delta_{jk'}\delta_{j'k}) \, r_j r'_{j'}
		\, \overline{ u_k(\bm{0})
						u_{k'}(\bm{r'-r})}
				\, \dd^3 \bm{r} \, \dd^3 \bm{r'}
\notag \\
	&= \frac{2}{3} \, \rho^2 k \int\!\!\!\!\int_{\varOmega^2}
		(r_j r'_j\delta_{kk'}-r_{k'} r'_k)
			\left( g(s)\delta_{kk'}
						+ [f(s)-g(s)] \frac{s_k s_{k'}}{s^2} \right)
				\, \dd^3 \bm{r} \, \dd^3 \bm{r'}
\notag \\
	&= \frac{2}{3} \, \rho^2 k \int\!\!\!\!\int_{\varOmega^2}
		\left( [f(s)+g(s)] \, r_k r'_k
				- [f(s)-g(s)] \frac{r'_k s_k r_{k'} s_{k'}}{s^2} \right)
				\, \dd^3 \bm{r} \, \dd^3 \bm{r'}
\notag \\
	&= \frac{2}{3} \, \rho^2 k \int\!\!\!\!\int_{\varOmega^2}
		\left( [f(s)+g(s)] \, \left[ x^2-s^2/4 \right]
			- [f(s)-g(s)] \left[ \frac{x_ks_kx_{k'}s_{k'}}{s^2}
								- s^2/4 \right] \right)
				\, \dd^3 \bm{s} \, \dd^3 \bm{x}
\notag \\ \label{eq:HH1}
	&= \frac{2}{3} \, \rho^2 k \int\!\!\!\!\int_{\varOmega^2}
		\left( f(s) \, \left[ x^2-\frac{(\bm{x \cdot s})^2}{s^2} \right]
			+ g(s) \, \left[ x^2+\frac{(\bm{x \cdot s})^2}{s^2}
							-s^2/2 \right] \right)
				\, \dd^3 \bm{s} \, \dd^3 \bm{x},
\end{align}
where homogeneity and isotropy have been taken into account, with $\bm{s}=\bm{r'-r}$, $\bm{x}=(\bm{r'+r})/2$, and $f$ and $g$ being the longitudinal and transverse correlation functions. The integration domain $\varOmega^2$ is represented in $\bm{x}$ and $\bm{s}$ variables as $x^2 \pm \bm{x \cdot s} + s^2/4 \leq R^2$, so using polar coordinates of $\bm{x}$ with respect to $\bm{s}$ (with $c=\bm{x \cdot s}/xs$), it is found
\begin{equation}
\int\!\!\!\!\int_{\varOmega^2}
		\dots \; \dd^3 \bm{s} \, \dd^3 \bm{x}
	= 8\pi^2 \int_0^{2R} \left(
		\int_0^{R-s/2} \!\! \int_{-1}^1 \dots \; x^2 \dd c \, \dd x
		+ \int_{R-s/2}^{\sqrt{R^2-s^2/4}}
			\!\! \int_{-(R^2-x^2-s^2/4)/xs}^{(R^2-x^2-s^2/4)/xs}
				\dots \; x^2 \dd c \, \dd x
		\right) s^2\dd s,
\end{equation}
and thus, after lengthy but straightforward calculations
\begin{multline}\label{eq:HH}
\overline{H^\varOmega_iH^\varOmega_i}
	= R^5 \frac{8\pi^2}{45} \, \rho^2 k \int_0^{2R} \Bigg[
			\left(8+\frac{9s}{2R}+\frac{3s^2}{4R^2}\right)
					\left(1-\frac{s}{2R}\right)^3 f(s)
\\
			+ \left(16-\frac{13s}{2R}-\frac{21s^2}{2R^2}
							-\frac{21s^3}{8R^3}\right)
					\left(1-\frac{s}{2R}\right)^2 g(s)
						\Bigg] \, s^2 \dd s.
\end{multline}

	Using the incompressibility condition $g(s)=f(s)+sf'(s)/2$ to eliminate $g(s)$ and performing integrations by parts on $s^nf'(s)$ terms to eliminate $f'(s)$, eventually yields
\begin{equation}\label{eq:HHi}
\overline{H^\varOmega_iH^\varOmega_i}
	= R^4 \frac{8\pi^2}{3} \, \rho^2 k \int_0^{2R}
		\left( 1-\frac{s^2}{2R^2} \right) \left( 1-\frac{s^2}{4R^2} \right)
					f(s) \, s^3 \dd s.
\end{equation}
It is to be noticed that the $R^5 \int fs^2\dd s$ and $f(2R)$ terms (the latter from partial integrations) \emph{both cancel exactly}---not just as $R\rightarrow\infty$,---leaving $R^4$ as leading term whenever $\int fs^3\dd s$ converges at $\infty$.
%
\subsection{3.~Torque variance on a sphere in HIT}

	The torque on a given spherical volume $\varOmega$ of radius $R$ around origin in a HIT field is given by Eq.~(\ref{eq:T}). Although it cancels on average, its average square norm does not and is given by
\begin{align}
\overline{T^\varOmega_iT^\varOmega_i}
	&= \int\!\!\!\!\int_{(\partial\varOmega)^2}
		\epsilon_{ijk} \epsilon_{ij'k'} r_j r'_{j'}
		\, \rho^2 \overline{ u_k(\bm{r}) u_l(\bm{r})
			u_{k'}(\bm{r'}) u_{l'}(\bm{r'})}
		\, \sigma_l \sigma'_{l'}
				\, \dd^2 \bm{r} \, \dd^2 \bm{r'}
\notag \\
	&= R^2 \rho^2 \int\!\!\!\!\int_{(\partial\varOmega)^2}
		(\sigma_j \sigma'_j\delta_{kk'}-\sigma_{k'} \sigma'_k)
		\, \overline{ u_k(\bm{0}) u_l(\bm{0})
				u_{k'}(\bm{r'-r})
						u_{l'}(\bm{r'-r})}
		\, \sigma_l \sigma'_{l'}
				\, \dd^2 \bm{r} \, \dd^2 \bm{r'},
\end{align}
where now $\bm{\hat\sigma}=\bm{r}/r=\bm{r}/R$ is the unit vector normal to the sphere surface $\partial\varOmega$. Here as for $\overline{H^\varOmega_iH^\varOmega_i}$ in the previous section, it is necessary to provide the general two-point quadruple velocity correlation tensor, which in HIT can be decomposed as
\begin{multline}\label{eq:4O2PCT}
\left(\frac{3}{2k}\right)^2 \overline{ u_k(\bm{0}) u_l(\bm{0})
						u_{k'}(\bm{s}) u_{l'}(\bm{s}) }
	= f_1(s) \frac{s_ks_ls_{k'}s_{l'}}{s^4}
	+ g_1(s) \left( \, \delta_{kl} \, \frac{s_{k'}s_{l'}}{s^2}
		+ \frac{s_ks_l}{s^2} \, \delta_{k'l'}
		- 2 \, \frac{s_ks_ls_{k'}s_{l'}}{s^4} \, \right)
\\
	+ g_2(s) \left( \, \frac{s_ks_{k'}}{s^2} \, \delta_{ll'}
		+ \frac{s_ks_{l'}}{s^2} \, \delta_{lk'}
		+  \frac{s_ls_{k'}}{s^2} \, \delta_{kl'}
		+  \frac{s_ls_{l'}}{s^2} \, \delta_{kk'}
		- 4 \, \frac{s_ks_ls_{k'}s_{l'}}{s^4} \, \right)
\\
	+ h_1(s) \, \left( \, \delta_{kl} \delta_{k'l'}
		- \frac{s_ks_l}{s^2} \, \delta_{k'l'}
		- \delta_{kl} \, \frac{s_{k'}s_{l'}}{s^2}
		+ \frac{s_ks_ls_{k'}s_{l'}}{s^2} \, \right)
\\
	+ h_2(s) \, \left( \, \delta_{kk'} \delta_{ll'}
		+ \delta_{kl'} \delta_{lk'}
		- \frac{s_ks_{k'}}{s^2} \, \delta_{ll'}
		- \frac{s_ks_{l'}}{s^2} \, \delta_{lk'}
		-  \frac{s_ls_{k'}}{s^2} \, \delta_{kl'}
		-  \frac{s_ls_{l'}}{s^2} \, \delta_{kk'}
\right. \\ \left.
		- 2 \delta_{kl} \delta_{k'l'}
		+ 2 \frac{s_ks_l}{s^2} \, \delta_{k'l'}
		+ 2 \delta_{kl} \, \frac{s_{k'}s_{l'}}{s^2} \, \right),
\end{multline}
corresponding to the five different symmetrized and orthogonalized correlations tensors along $zz|zz$, $zz|xx$, $zx|zx$, $xx|yy$, and $xy|xy$ when $\bm{s}$ is along $z$. The correlation functions have the following elementary properties
\begin{subequations}\begin{align}
g_2(0) &= g_1(0),
&
\lim_{s \rightarrow \infty} f_1(s), g_1(s), h_1(s) &= 1,
\\
h_2(0) &= h_1(0),
&
\lim_{s \rightarrow \infty} g_2(s), h_2(s) &= 0,
\\
f_1(s), g_1(s), h_1(s) &> 0.
\end{align}\end{subequations}
Taking into account that $\bm{\hat\sigma}$ and $\bm{\hat\sigma'}$ are unit vectors, that $\bm{s}=\bm{r'}-\bm{r}=R(\bm{\hat\sigma'-\hat\sigma})$, and defining $\bm{\hat\sigma\cdot\hat\sigma'}=\cos\theta=1-s^2/(2R^2)$, the various contractions of the elementary tensors yield
\begin{subequations}\begin{align}
(\sigma_j \sigma'_j\delta_{kk'}-\sigma_{k'} \sigma'_k)
		\, \frac{s_ks_ls_{k'}s_{l'}}{s^4} \, \sigma_l \sigma'_{l'}
	&= \frac{c^2-1}{4},
\\
(\sigma_j \sigma'_j\delta_{kk'}-\sigma_{k'} \sigma'_k)
		\left( \, \delta_{kl} \, \frac{s_{k'}s_{l'}}{s^2}
		+ \frac{s_ks_l}{s^2} \, \delta_{k'l'} \, \right)
			\, \sigma_l \sigma'_{l'}
	&= 0,
\\
(\sigma_j \sigma'_j\delta_{kk'}-\sigma_{k'} \sigma'_k)
		\left( \, \frac{s_ks_{k'}}{s^2} \, \delta_{ll'}
		+ \frac{s_ks_{l'}}{s^2} \, \delta_{lk'}
		+  \frac{s_ls_{k'}}{s^2} \, \delta_{kl'}
		+  \frac{s_ls_{l'}}{s^2} \, \delta_{kk'} \, \right)
			\, \sigma_l \sigma'_{l'}
	&= \frac{5c^2-c-2}{2},
\\
(\sigma_j \sigma'_j\delta_{kk'}-\sigma_{k'} \sigma'_k)
		\, \delta_{kl} \delta_{k'l'} \sigma_l \sigma'_{l'}
	&= 0,
\\
(\sigma_j \sigma'_j\delta_{kk'}-\sigma_{k'} \sigma'_k)
		\left( \, \delta_{kk'} \delta_{ll'}
		+ \delta_{kl'} \delta_{lk'} \, \right)
			\, \sigma_l \sigma'_{l'}
	&= 3c^2-1.
\end{align}\end{subequations}
Therefore
\begin{equation} \label{eq:TTs}
\overline{T^\varOmega_iT^\varOmega_i}
	= R^2 \left(\frac{2 \rho k}{3}\right)^2
				\int\!\!\!\!\int_{(\partial\varOmega)^2}
	\left[\, \frac{c^2-1}{4}\,\Big( f_1(s)-2g_1(s)+h_1(s) \Big)
			+  \frac{3c^2-5c}{2}\,g_2(s) +  \frac{c^2+5c}{2}\,h_2(s)
			\,\right] \; \dd^2 \bm{r} \dd^2 \bm{r'}.
\end{equation}
Using polar coordinates relative to $\bm{r}$ for instance, so $\bm{r \cdot r'}=r^2\cos\theta$, and substituting $\cos\theta=1-s^2/(2R^2)$
\begin{multline}\label{eq:TT}
\overline{T^\varOmega_iT^\varOmega_i}
	= 2 R^4 \left(\frac{2\pi\rho k}{3}\right)^2 \int_0^{2R} \bigg\{
		\left( 1-\frac{s^2}{2R^2} \right)
			\left[\, \left( 12-\frac{s^2}{R^2} \right) h_2(s)
				- \left( 4+3\,\frac{s^2}{R^2} \right) g_2(s) \,\right]
\\
		+ \frac{s^2}{R^2} \left( 1-\frac{s^2}{4R^2} \right)
			\left[\, f_1(s)-2g_1(s)+h_1(s) \,\right]
	\bigg\} \, s \; \dd s.
\end{multline}

	Here again the incompressibility conditions could produce some simplifications. It is necessary however, to first obtain these conditions on the two-point quadruple correlation tensor in Eq.~(\ref{eq:4O2PCT}). The tensor is purely rotational in any of its components, and by symmetry, a zero divergence with respect to the first component suffices to ensure incompressibility. Now, from its general form in Eq.~(\ref{eq:4O2PCT}), substituted by monomial tensors $T^\alpha_{klk'l'}$ and functions $f_\alpha$ as combinations of $f_1$ to $h_2$, and using $s_{,k}=s_k/s$, the incompressibility condition can be written
\begin{equation}
\bigg( \sum_\alpha f_\alpha(s) T^\alpha_{klk'l'}(s) \bigg)_{,k} B_{lk'l'}
	= \sum_\alpha f_\alpha'(s) \, \frac{s_k}{s} \, T^\alpha_{klk'l'} B_{lk'l'}
		+ f_\alpha(s) T^\alpha_{klk'l',k} B_{lk'l'} = 0,
\end{equation}
where $B_{lk'l'}$ is any arbitrary tensor. The decomposition of the third-order tensors $B_{lk'l'}$ in irreducible representations taking into account the symmetry properties shows that three scalar conditions are eventually obtained, corresponding to the three following orthogonal tensors $B_{lk'l'}$
\begin{align}
\frac{s_ls_{k'}s_{l'}}{s^3},
&& \frac{s_l}{s} \, \delta_{k'l'} - \frac{s_ls_{k'}s_{l'}}{s^3},
&& \delta_{ll'} \frac{s_{k'}}{s} - \frac{s_ls_{k'}s_{l'}}{s^3},
\end{align}
or equivalently, to the three independent tensors
\begin{align}
\frac{s_ls_{k'}s_{l'}}{s^3},
&& \frac{s_l}{s} \, \delta_{k'l'},
&& \delta_{ll'} \frac{s_{k'}}{s}.
\end{align}
The table of the different contractions $(s_k/s)T^\alpha_{klk'l'} B_{lk'l'}$ and $T^\alpha_{klk'l',k} B_{lk'l'}$ is thus obtained:
\begin{equation*}
\begin{array}{*{4}{>{\displaystyle~~~~}c<{~~}>{\displaystyle~~}c<{~~~~}}}
T^\alpha_{klk'l'} & T^\alpha_{klk'l',k}
& \multicolumn{2}{|c}{\displaystyle\frac{s_ls_{k'}s_{l'}}{s^3}}
& \multicolumn{2}{|c}{\displaystyle\frac{s_l}{s} \, \delta_{k'l'}}
& \multicolumn{2}{|c}{\displaystyle\delta_{ll'} \frac{s_{k'}}{s}}
\\[.7em]\hline\rule[0em]{0em}{1.7em}
	\frac{s_ks_ls_{k'}s_{l'}}{s^4}
	& 2 \, \frac{s_ls_{k'}s_{l'}}{s^4}
	& 1 & \frac{2}{s} & 1 & \frac{2}{s} & 1 & \frac{2}{s}
\\[.7em]
	\delta_{kl} \frac{s_{k'}s_{l'}}{s^2}
	& \delta_{lk'} \frac{s_{l'}}{s^2} + \delta_{ll'} \frac{s_{k'}}{s^2}
		- 2 \, \frac{s_ls_{k'}s_{l'}}{s^4}
	& 1 & 0 & 1 & 0 & 1 & \frac{2}{s}
\\[.7em]
	\frac{s_ks_l}{s^2} \, \delta_{k'l'}
	& 2 \, \frac{s_l}{s^2} \, \delta_{k'l'}
	& 1 & \frac{2}{s} & 3 & \frac{6}{s} & 1 & \frac{2}{s}
\\[.7em]
	\frac{s_ks_{k'}}{s^2} \, \delta_{ll'}
	& 2 \, \frac{s_{k'}}{s^2} \, \delta_{ll'}
	& 1 & \frac{2}{s} & 1 & \frac{2}{s} & 3 & \frac{6}{s}
\\[.7em]
	\frac{s_ks_{l'}}{s^2} \, \delta_{lk'}
	& 2 \, \frac{s_{l'}}{s^2} \, \delta_{lk'}
	& 1 & \frac{2}{s} & 1 & \frac{2}{s} & 1 & \frac{2}{s}
\\[.7em]
	\frac{s_ls_{k'}}{s^2} \delta_{kl'}
	& \frac{s_{k'}}{s^2} \delta_{ll'} + \frac{s_l}{s^2} \, \delta_{k'l'}
		- 2 \, \frac{s_ls_{k'}s_{l'}}{s^4}
	& 1 & 0 & 1 & \frac{2}{s} & 1 & \frac{2}{s}
\\[.7em]
	\frac{s_ls_{l'}}{s^2} \delta_{kk'}
	& \frac{s_{l'}}{s^2} \delta_{lk'} + \frac{s_l}{s^2} \, \delta_{k'l'}
		- 2 \, \frac{s_ls_{k'}s_{l'}}{s^4}
	& 1 & 0 & 1 & \frac{2}{s} & 1 & 0
\\[.7em]
	\delta_{kl} \delta_{k'l'}
	& 0
	& 1 & 0 & 3 & 0 & 1 & 0
\\[.7em]
	\delta_{kk'} \delta_{ll'}
	& 0
	& 1 & 0 & 1 & 0 & 3 & 0
\\[.7em]
	\delta_{kl'} \delta_{k'l}
	& 0
	& 1 & 0 & 1 & 0 & 1 & 0
\\[.7em]\hline
\end{array}
\end{equation*}
Hence the three conditions
\begin{subequations}\begin{align}
& sf_1'/2 + f_1 - g_1 - 2g_2 = 0,
\\
& sg_1'/2 + g_1 + g_2 - h_1 + h_2 = 0,
\\
& sg_2' + 3g_2 + g_1 - h_1 - h_2 = 0.
\end{align}\end{subequations}
To reduce the $R^4$ scaling of $\overline{T^\varOmega_iT^\varOmega_i}$ in Eq.~(\ref{eq:TT}), the leading integrand term proportional to $(3h_2-g_2)s\dd s$ should be canceled by combining the incompressibility constraints and performing integrations by parts. Taking coefficients $\alpha$, $\beta$, and $\gamma$ for the three constraints, this leading term is written as:
\begin{multline}
\textstyle
\big[ (3h_2-g_2)
	+ \alpha (sf_1'/2 + f_1 - g_1 - 2g_2)
	+ \beta (sg_1'/2 + g_1 + g_2 - h_1 + h_2)
	+ \gamma (sg_2' + 3g_2 + g_1 - h_1 - h_2) \big]s\dd s
\\
= \big[ -(\alpha-\gamma) g_1
	- (2\alpha-\beta-\gamma+1) g_2
	- (\beta+\gamma) h_1
	+ (\beta-\gamma+3) h_2 \big] s \dd s,
+ \dd \big[ s^2 ( \alpha f_1 + \beta g_1 + \gamma g_2 ) \big].
\end{multline}
In general, no combinations of $\alpha$, $\beta$, and $\gamma$ can be found to simultaneously cancel the four integrands on the right hand side. The scaling in $R^4$ as given in Eq.~(\ref{eq:TT}) is thus completely general, and the incompressibility conditions just provide the possibility of expressing the leading term as a function of only one of the four correlation functions $g_1$, $g_2$, $h_1$, or $h_2$---a substitution of minor interest.
%
\subsection{4.~Scalings of $\overline{H^\varOmega_zH^\varOmega_z}$ and $\overline{T^\varOmega_zT^\varOmega_z}$ in the basic $C_\infty$ configurations}
\end{widetext}

\paragraph{1.~Generic scaling approach.}

	As mentioned in the main text, the impulsive picture of the velocity fluctuations can accurately provide scaling laws for $\overline{H^\varOmega_zH^\varOmega_z}$ and $\overline{T^\varOmega_zT^\varOmega_z}$ in HIT, and in layer and tube of turbulence. This is ensured by the basic property of the Saffman procedure which produces kinematically consistent velocity fields out from arbitrary impulse fields, while preserving at the same time the angular momentum $H^\varOmega_z$.

	The case of HIT is represented in Fig.~1. Within structures of typical size $\lambda$, the velocity field is modeled as fully correlated. In the Saffman case, this produces contributions from each structure to the double integral in $\overline{H^\varOmega_zH^\varOmega_z}$ scaling as $\sim kR^2\lambda^6$. The scaling is reduced to $\sim k\lambda^2\lambda^6$ in the Batchelor case where motions are purely rotational, since opposite contributions from within a given structure cancel in first order of $R$. Between structures, the velocity field is now modeled as fully uncorrelated: the double sum on $\varOmega$ in $\overline{H^\varOmega_zH^\varOmega_z}$ multiplies the structure contribution by $\sim(R/\lambda)^3$, the typical number of structures in volume $\varOmega$.

	This general approach to estimating scalings is simply adapted for $\overline{T^\varOmega_zT^\varOmega_z}$ and for the layer and tube of turbulence. The various intermediate contributions and final results are collected in Table~\ref{tab:HT}, but some peculiarities require further comments as given below.

\paragraph{2.~Relationship with rigorous calculations for HIT.}

	As already mentioned in the main text, the rigorous calculation in §2 of $\overline{(H^\varOmega_z)^2}$ for HIT gives Eq.~(\ref{eq:HHi}), with a leading contribution in $R^4$ under incompressibility constraints which are not fulfilled by the impulse fields considered here. A $R^5$ behavior is actually found in the general case in Eq.~(\ref{eq:HH}), compatible with the Saffman case. A $R^3$ behavior could be recovered with a specific oscillating profile of the correlation functions $f$ and $g$, as would be expected in the Batchelor case.

\paragraph{3.~Scaling of $\overline{T^\varOmega_zT^\varOmega_z}$.}

	The estimation of $\overline{T^\varOmega_zT^\varOmega_z}$ follows the same general lines as for $\overline{H^\varOmega_zH^\varOmega_z}$. For the Saffman case of HIT in Fig.~1a, contributions of each structure to the double surface integral in $\overline{T^\varOmega_zT^\varOmega_z}$ scale as $\sim k^2R^2\lambda^4$. Now, at variance with the estimation of $\overline{H^\varOmega_zH^\varOmega_z}$ given above, this scaling is conserved in the Batchelor case of Fig.~1b and must \emph{not} be reduced to $\sim k^2\lambda^2\lambda^4$: opposite contributions from within a given structure are not integrated over its full volume but on a randomly oriented slice only, and therefore the cancellation of the torque in first order of $R$ is incomplete. Finally, in all cases of impulse fields, the double sum on $\partial\varOmega$ in $\overline{T^\varOmega_zT^\varOmega_z}$ multiplies the structure contribution by $\sim(R/\lambda)^2$, the typical number of structures intersecting surface $\partial\varOmega$. Therefore, the $\overline{T^\varOmega_zT^\varOmega_z}$ scaling is independent of the type of the correlations in the impulse field, and sets the maximum scaling exponent of $\overline{H^\varOmega_zH^\varOmega_z}$ compatible with asymptotic invariance at $R\rightarrow\infty$. As already mentioned in the main text, although the Batchelor impulse field does not produce an invariant, it should still be considered here for its relevance in experiments and simulations, where physical boundaries to the flows are always present at some large scale. This important scaling property of $\overline{T^\varOmega_zT^\varOmega_z}$ is also found in the layer and tube of turbulence, and Table~\ref{tab:HT} thus lists the corresponding invariants of maximum exponents.

\paragraph{4.~Basic scalings in configurations other than HIT.}

	The major difference between the different flows considered here comes from the double integral over $\varOmega$ in $\overline{H^\varOmega_zH^\varOmega_z}$: proportional to $(R/\lambda)^3$, $(R/\lambda)^2$, and $R/\lambda$ for HIT, layer, and tube respectively. An interesting feature is also to be noticed on the point contribution to $\overline{H^\varOmega_zH^\varOmega_z}$ for the tube of turbulence: the typical distance to the $z$ axis of any given point in the turbulent zone is $\lambda$, instead of $R$ for the HIT and layer of turbulence. Accordingly, all the possible types of impulse fields---here the Saffman and Batchelor cases---yield identical $\overline{H^\varOmega_zH^\varOmega_z}$ scalings. The evolution of the tube of turbulence is thus more universal as it is independent of the long-range correlations in the initial conditions.

\paragraph{5.~Spinning tube of turbulence.}

	The integration volumes $\varOmega$ used for deriving asymptotic invariants are all of $C_\infty$ symmetry and are thus also invariant by rotations. It is therefore natural to consider non-vanishing mean velocity fields producing global spinning of the turbulent zones. However, aside from the spot of turbulence examined below, the tube of turbulence is the only flow configuration where a homogeneous spinning will not produce divergent centrifugal forces. A spinning tube of turbulence could see its turbulence relaxation controlled by the conservation of the angular momentum itself, $\overline{H^\varOmega_z}$ instead of $\overline{H^\varOmega_zH^\varOmega_z}$. This somewhat different situation implies that (i) the mean velocity field is non zero in the turbulent zone, (ii) the flow evolution produces an equilibration between the mean kinetic energy contained in the angular momentum and the turbulent kinetic energy---a reasonable assumption,---and (iii) the mean angular momentum dominates its variance, otherwise the evolution would be constrained by the large scale invariance of $\overline{H^\varOmega_zH^\varOmega_z}$. The scaling in Table~\ref{tab:HT} is then found, given as a function of $\overline{H^\varOmega_z}\times\overline{H^\varOmega_z}$ for consistency with the other results.

\paragraph{6.~Spinning spot of turbulence.}

	Because the extension of a spot of turbulence is finite in all directions, it is not possible to predict its evolution from angular momentum variance at large scales. More directly, the angular momentum itself is invariant for the volume $\varOmega$ in Fig.~3. As for the spinning tube of turbulence, this implies (i) a non-zero average velocity field, (ii) an equilibration between the mean and turbulent kinetic energies, and (iii) a non-negligible angular momentum, otherwise the evolution of the corresponding quadrupolar---or higher multipolar---spot would not be constrained by any obvious invariant. The scaling in Table~\ref{tab:HT} is then found, again as a function of $\overline{H^\varOmega_z}\times\overline{H^\varOmega_z}$.

\paragraph{7.~Axes other than $z$ for angular momentum conservation.}

	In principle, for the geometries other than HIT of Fig.~3, one could consider the asymptotic invariance of $\overline{H^\varOmega_xH^\varOmega_x}$---not $\overline{H^\varOmega_zH^\varOmega_z}$---for a \emph{spherical} volume $\varOmega$, instead of slab and rod shaped volumes. Pressure forces would still not contribute to the torque, but the fringe field of velocity fluctuations in the laminar regions would contribute substantially and in a complex way to both the angular momentum and torque, as the laminar regions would represent the largest volume fractions of the sphere. The corresponding invariance conditions could then actually \emph{not} constrain the evolution of the turbulence zone, although they would be compatible with the present findings based on $\overline{H^\varOmega_zH^\varOmega_z}$.

\paragraph{8.~Contributions to $\overline{H^\varOmega_zH^\varOmega_z}$ from laminar zones of $\varOmega$.}

	As shown in Fig.~3, the domains $\varOmega$ considered for angular momentum conservation at large scales extend somewhat beyond the turbulent zones. In these laminar regions, velocity fluctuations exist due to the long range influence of the motions in the turbulent zone, although they decay quickly with distance. These velocity fluctuations \emph{could} contribute to $H^\varOmega_z$, despite the fact that $H^\varOmega_z$ is preserved by Saffman's projection procedure in Eq.~(6): indeed, Saffman's projection only deals with a pressure gradient correction to the impulse field, but does not involve transport which is the source of torque in Eq.~(\ref{eq:T}). Now, as velocity in the laminar region can be described by a potential flow whereby $\bm{u}=\bm{\nabla}\phi$, the contribution to $H^\varOmega_z$ of a given volume $\varOmega_\ell$ in the laminar zone of $\varOmega$ with the \emph{same} $C_\infty$ symmetry is
\begin{multline}
H^{\varOmega_\ell}_z
	= \int_{\varOmega_\ell} \epsilon_{zij} r_i \rho u_j \,\dd^3 \bm{r}
\\
	= \int_{\varOmega_\ell} \epsilon_{zij} r_i \rho \phi_{,j} \,\dd^3 \bm{r}
	= \int_{\varOmega_\ell} (\epsilon_{zij} r_i \rho \phi)_{,j} \,\dd^3 \bm{r}
\\
	= \int_{\partial\varOmega_\ell}
			\epsilon_{zij} r_i \rho \phi \, \sigma_j \,\dd^2 \bm{r}
	= 0.
\end{multline}
The final cancellation comes from the symmetry of $\varOmega_\ell$, so $\bm{r\times\hat{\sigma}}$ has no component along $z$ ($\bm{\hat{\sigma}}$ is the unit vector normal to the surface $\partial\varOmega_\ell$). It also assumes that $\varOmega_\ell$ is simply connected, or is just connected but does not carry any net circulation of $\bm{u}$ around singularities (as could be the case for the tube of turbulence).

\paragraph{9.~Contributions to $\overline{T^\varOmega_zT^\varOmega_z}$ from laminar zones of $\partial\varOmega$.}
\begin{figure}
\includegraphics[width=2.5in]{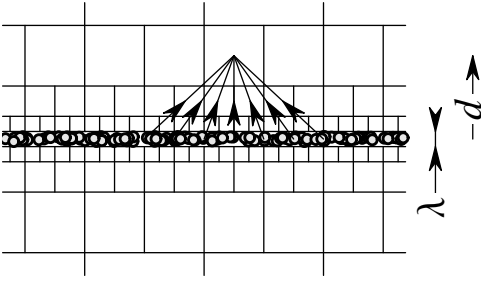}
\caption{\label{fig:Evan}Schematic representation of the build up of velocity fluctuations and correlations in the laminar surroundings of a turbulent zone (layer or tube). Arrows represent the potential field contributions of turbulent structures affecting a given point of the laminar zone. Squares represent typical domains of correlated fluctuations.}
\end{figure}

	As already mentioned, angular momentum cancels in the laminar regions but can be transported across these regions to and from the turbulent zone. It is thus necessary to evaluate the magnitude of the torque contributions on the side surfaces of the domains $\varOmega$ represented in Fig.~3. As with the contribution from the turbulent zone estimated above, the double integral in $\overline{T^\varOmega_zT^\varOmega_z}$ will be defined by the typical velocity fluctuations at each point, summed over the surface in correlated or uncorrelated ways.

	Let $d$ be the distance of the surface of $\varOmega$ to the turbulent zone. The velocity fluctuations at distance $d$ are given by the potential field produced by the turbulent zone in the laminar region, which scales as $(\lambda/d)^3\sqrt{k}$ or $(\lambda/d)^4\sqrt{k}$ for a single structure of dipolar or quadrupolar character respectively (Saffman or Batchelor cases). As sketched in Fig.~\ref{fig:Evan}, the effective field at distance $d$ is the superposition of uncorrelated contributions coming from structures within a typical radius $d$: this yields the various point contributions as listed in Table~\ref{tab:HT}. Since at distance $d$ the field is produced by turbulent structures within a typical radius $d$, velocity fluctuations are also correlated over a typical distance $d$: this yields the various surface integration factors as listed in Table~\ref{tab:HT}. Combining the different factors, the relative magnitudes of the laminar contributions to $\overline{T^\varOmega_zT^\varOmega_z}$ are then found as listed in Table~\ref{tab:HT}.

	In order to preserve the same invariance conditions as defined by the scalings in $R$ of $\overline{H^\varOmega_zH^\varOmega_z}$ and $\overline{T^\varOmega_zT^\varOmega_z}$, it is necessary to make the relative laminar contribution to $\overline{T^\varOmega_zT^\varOmega_z}$ decrease with $R$. Upon inspection of results in Table~\ref{tab:HT}, this is always possible in all cases by making $d$ vary with $R$, for instance as $d \propto R^\alpha\lambda^{(1-\alpha)}$ with $1/5\leq\alpha<1$. This justifies, a posteriori, the requirement that $\varOmega$ extends somewhat into the laminar region.

\paragraph{10.~Contributions to $\overline{T^\varOmega_zT^\varOmega_z}$ from laminar zones of $\partial\varOmega$ in spinning tube and spot.}

	The derivation of the previous paragraph must be somewhat adapted to the cases of spinning tube and spot of turbulence. The fluctuations in the surrounding laminar fluid due to the rotation of the turbulent zone come from the irregular shape of the fluid domain entrained by the rotation. The potential flow around a spinning spot of turbulence would then produce velocity fluctuations decaying at least as $(\lambda/d)^4$, as for a Batchelor structure, faster decays being possible for shapes of rotating fluid domains of higher multipolar structure. Hence the results for the spot of turbulence listed in Table~\ref{tab:HT}. It should be noticed that, for the spot, the torque on the surface $\partial\varOmega_\ell$ in the laminar domain actually represents the full torque, and the asymptotic invariance of $\overline{H^\varOmega_z}\times\overline{H^\varOmega_z}$ is now obtained for $d\rightarrow\infty$. The results for the spinning tube of turbulence are directly adapted from the spot case, assuming uncorrelated phases and shapes of the spinning sections of typical size $\lambda$ over length $R$.

\begin{turnpage}
\begin{table}
\caption{\label{tab:HT}Magnitudes of leading terms in $R$ to $\overline{H^\varOmega_zH^\varOmega_{z}}$ and $\overline{T^\varOmega_zT^\varOmega_{z}}$ for the various flow configurations and impulse field types considered here, and corresponding torque contributions from laminar zones (TZ = turbulent zone, LS = laminar surroundings).}
\begin{ruledtabular}
\begin{tabular}{l*{8}{c}}
Geometry
	& \multicolumn{2}{c}{HIT}
	& \multicolumn{2}{c}{Layer (IRM)}
	& \multicolumn{3}{c}{Tube}
	& \multicolumn{1}{c}{Spot}
		\\ \cline{2-3} \cline{4-5} \cline{6-8} \cline{9-9}
\rule[0em]{0em}{1.1em}%
Invariant type
	& Saffman & Batchelor
	& Saffman & Batchelor
	& Saffman & Batchelor & Spin
	& Spin \\ \hline
\rule[0em]{0em}{1.1em}%
Point contribution to $\overline{H^\varOmega_zH^\varOmega_{z}}$ in TZ
	& $kR^2$ & $k\lambda^2$
	& $kR^2$ & $k\lambda^2$
	& $k\lambda^2$ & $k\lambda^2$ & $k\lambda^2$
	& $k\lambda^2$ \\
Sum over volume in TZ
	& \multicolumn{2}{c}{$\times\lambda^6\times(R/\lambda)^3$}
	& \multicolumn{2}{c}{$\times\lambda^6\times(R/\lambda)^2$}
	& \multicolumn{2}{c}{$\times\lambda^6\times(R/\lambda)$}
	& $\times\lambda^6\times(R/\lambda)^2$
	& $\times\lambda^6$ \\
$\overline{H^\varOmega_zH^\varOmega_{z}}$ scaling in TZ
	& $R^5k\lambda^3$ & $R^3k\lambda^5$
	& $R^4k\lambda^4$ & $R^2k\lambda^6$
	& $Rk\lambda^7$ & $Rk\lambda^7$ & $R^2k\lambda^6$
	& $k\lambda^8$ \\
Point contribution to $\overline{T^\varOmega_zT^\varOmega_{z}}$ in TZ
	& \multicolumn{2}{c}{$k^2R^2$}
	& \multicolumn{2}{c}{$k^2R^2$}
	& \multicolumn{2}{c}{$k^2\lambda^2$}
	& $k^2\lambda^2$
	& -- \\
Sum over surface in TZ
	& \multicolumn{2}{c}{$\times\lambda^4\times(R/\lambda)^2$}
	& \multicolumn{2}{c}{$\times\lambda^4\times(R/\lambda)$}
	& \multicolumn{2}{c}{$\times\lambda^4$}
	& $\times\lambda^4$
	& -- \\
$\overline{T^\varOmega_zT^\varOmega_{z}}$ scaling in TZ
	& \multicolumn{2}{c}{$R^4k^2\lambda^2$}
	& \multicolumn{2}{c}{$R^3k^2\lambda^3$}
	& \multicolumn{2}{c}{$k^2\lambda^6$}
	& $k^2\lambda^6$
	& $(\lambda/d)^{10}k^2\lambda^6$ \\ \hline
\rule[0em]{0em}{1.1em}%
Point contribution to $\overline{T^\varOmega_zT^\varOmega_{z}}$ in LS
	& -- & --
	& $(\lambda/d)^8k^2R^2$ & $(\lambda/d)^{12}k^2R^2$
	& $(\lambda/d)^{10}k^2d^2$ & $(\lambda/d)^{14}k^2d^2$
	& $(\lambda/d)^{14}k^2d^2$
	& $(\lambda/d)^{16}k^2d^2$ \\
Sum over surface in LS
	& \multicolumn{2}{c}{--}
	& \multicolumn{2}{c}{$\times d^4\times(R/d)^2$}
	& \multicolumn{2}{c}{$\times d^4\times(R/d)$}
	& $\times d^4\times(R/d)$
	& $\times d^4$ \\
$\overline{T^\varOmega_zT^\varOmega_{z}}$: scaling of LS to TZ contributions ratio
	& -- & --
	& $\lambda^5R/d^6$ & $\lambda^9R/d^{10}$
	& $\lambda^4R/d^5$ & $\lambda^8R/d^9$
	& $\lambda^8R/d^9$
	& --
\end{tabular}
\end{ruledtabular}
\end{table}
\end{turnpage}
\end{document}